\documentstyle[12pt,epsf,epsfig]{article}
\newcount\timecount
\newcount\hours \newcount\minutes  \newcount\temp \newcount\pmhours

\hours = \time
\divide\hours by 60
\temp = \hours
\multiply\temp by 60
\minutes = \time
\advance\minutes by -\temp
\def\hour{\the\hours}
\def\minute{\ifnum\minutes<10 0\the\minutes
            \else\the\minutes\fi}
\def\clock{
\ifnum\hours=0 12:\minute\ AM
\else\ifnum\hours<12 \hour:\minute\ AM
      \else\ifnum\hours=12 12:\minute\ PM
            \else\ifnum\hours>12
                 \pmhours=\hours
                 \advance\pmhours by -12
                 \the\pmhours:\minute\ PM
                 \fi
            \fi
      \fi
\fi
}

\def\monthname{\relax\ifcase\month 0/\or January\or February\or
   March\or April\or May\or June\or July\or August\or September\or
   October\or November\or December\else\number\month/\fi}

\def\bold#1{\setbox0=\hbox{$#1$}%
     \kern-.025em\copy0\kern-\wd0
     \kern.05em\copy0\kern-\wd0
     \kern-.025em\raise.0433em\box0 }

\def\gappeq{\mathrel{\rlap {\raise.5ex\hbox{$>$}}
{\lower.5ex\hbox{$\sim$}}}}

\def\lappeq{\mathrel{\rlap{\raise.5ex\hbox{$<$}}
{\lower.5ex\hbox{$\sim$}}}}

\def\ga{\mathrel{\raise.3ex\hbox{$>$\kern-.75em\lower1ex\hbox{$\sim$}}}}
\def\la{\mathrel{\raise.3ex\hbox{$<$\kern-.75em\lower1ex\hbox{$\sim$}}}}
\def\gev{{\rm \, Ge\kern-0.125em V}}
\def\tev{{\rm \, Te\kern-0.125em V}}
\def\beq{\begin{equation}}
\def\eeq{\end{equation}}

\newcommand\iso[2]{\mbox{${}^{#2}${\rm #1}}}
\newcommand\he[1]{\iso{He}{#1}}
\newcommand\li[1]{\iso{Li}{#1}}

\newcommand\pref[1]{(\ref{#1})}
\newcommand\omb{\Omega_{\rm B}}
\newcommand\nnu{\mbox{$N_{\rm \nu,eff}$}}

\newcommand\like{{\cal L}}
\newcommand\etal{{\it et al.~}}

\begin{document}
\begin{titlepage}
\pagestyle{empty}
\baselineskip=21pt
\rightline{astro-ph/0105397}
\rightline{CERN--TH/2001-129}
\rightline{UMN--TH--2005/01, TPI--MINN--01/21}
\vskip 0.35in
\begin{center}
{\large{\bf Primordial Nucleosynthesis with CMB Inputs: \\
Probing the Early Universe and Light Element Astrophysics }}
\end{center}
\begin{center}
\vskip 0.05in
{{\bf Richard H. Cyburt}$^1$, {\bf Brian D. Fields}$^2$, and
{\bf Keith A.~Olive}$^{3,4}$
\vskip 0.05in
{\it
$^1${Department of Physics\\ University of Illinois, Urbana, IL 61801,
USA}
\\
$^2${Center for Theoretical Astrophysics, Department of
Astronomy\\  University of Illinois, Urbana, IL 61801, USA}
\\
$^3${TH Division, CERN, Geneva, Switzerland}\\
$^4${Theoretical Physics Institute, School of Physics and
Astronomy,\\ University of Minnesota, Minneapolis, MN 55455,
USA}\\ }}
\vskip 0.35in
{\bf Abstract}
\end{center}
\baselineskip=18pt \noindent

Cosmic microwave background (CMB) determinations of the baryon-to-photon
ratio $\eta \propto \Omega_{\rm baryon} h^2$ will remove the last
free parameter from (standard) big bang nucleosynthesis (BBN)
calculations.  This will make BBN a much sharper probe
of early universe physics, for example, greatly refining the BBN
measurement of the effective number of light neutrino species, $\nnu$.  We
show how the CMB can improve this limit, given current light element
data.   Moreover, it will become possible to constrain $\nnu$
independent of \he4, by using other elements, notably
deuterium; 
this will allow for sharper limits and tests of systematics.
For example, a
3\% measurement of $\eta$, together with a 10\% (3\%) 
measurement of
primordial D/H, can measure $\nnu$ to a $95\%$ confidence level of
$\sigma_{95\%}(\nnu) = 1.8$ (1.0)
if \ $\eta \sim 6.0\times 10^{-10}$. 
If instead, one adopts the standard model value $\nnu=3$, then one can
use $\eta$ (and its uncertainty) from the CMB to make accurate predictions
for the primordial abundances. These determinations can in turn  become
key inputs in the nucleosynthesis history (chemical
evolution) of galaxies thereby placing constraints on such models.

\vfill
\vskip 0.15in
\leftline{May 2001}
\end{titlepage}
\baselineskip=18pt

\section{Introduction}

Cosmology is currently undergoing a revolution
spurred by a host of new precision observations.
A key element in this revolution is the measurement of
the anisotropy in the cosmic microwave background (CMB)
at small angular scales \cite{boom} - \cite{newboom}. 
In principle, an accurate determination of the CMB anisotropy allows
for the precision measurement of cosmological parameters, including
a very accurate determination of the baryon density
$\rho_{\rm B} \propto \omb h^2$.  
Because the present mean temperature of the CMB 
is extremely well-measured, one can then infer
the baryon-to-photon ratio $\eta = n_{\rm B}/n_\gamma$, 
via $\eta_{10} = \eta\times\! 10^{10} = 274 \omb h^2$.

To date, big bang nucleosynthesis (BBN) provides
the best measure of $\eta$, as this
is the only free parameter of standard BBN
(assuming the number of neutrino species $\nnu = 3$,
as in the standard electroweak model; see
below). The CMB anisotropies thus {\em independently test}
the BBN prediction \cite{cmbtest}.  
Initial measurements of the CMB anisotropy
already allow for the first tests of CMB-BBN consistency.
At present, the predicted BBN
baryon densities agree to an uncanny level with the most recent
CMB results \cite{dasi,newboom}.   The recent result from DASI \cite{dasi}
indicates that $\Omega_B h^2 = 0.022^{+0.004}_{-0.003}$, while that of 
BOOMERanG-98  \cite{newboom}, 
$\Omega_B h^2 = 0.021^{+0.004}_{-0.003}$ (using 1$\sigma$
errors) which should be  compared to the BBN predictions, $\Omega_B h^2 = 0.021$
with a 95\% CL range of 0.018 -- 0.027,
based only on D/H in high redshift quasar absorption systems \cite{omear}. These
determinations are higher than the value $\Omega_B h^2 = 0.009$ (0.006 -- 0.017 95\%
CL) based on \he4 and \li7 \cite{cfo}. However, the measurements of the Cosmic
Background Imager (CBI; Padin et al.\ \cite{cbi}) 
at smaller angular scales (higher multipoles) agree with lower BBN predictions
and claims a maximum likelihood value for $\Omega_B h^2 = 0.009$ (albeit with a
large uncertainty). 
We also note that in the DASI analysis \cite{dasi},
values of $\Omega_B h^2 < 0.01$ were not considered, and therefore we
consider their result an upper limit to the baryon density.

In this paper
we anticipate the impact on BBN of
future high-precision CMB experiments.
We begin with a summary (\S \ref{sect:formal}) 
of BBN analysis.
In \S \ref{sect:test} we examine the test of cosmology 
which will come from
comparing the BBN and CMB determinations of the cosmic baryon 
density.
In \S \ref{sect:BBN+CMB} we describe and quantify the enhanced
ability to probe the early universe, 
and quantify the precision with which primordial
abundances can be predicted and thereby constrain 
various astrophysical processes.
The impact of improvements in the observed abundances and
theoretical inputs are discussed in \S \ref{sect:improve},
and discussion and conclusions appear in \S \ref{sect:dis}.

\section{Formalism and Strategy}
\label{sect:formal}

As is well known, BBN is sensitive to physics at the epoch
$t \sim 1$ sec, $T \sim 1$ MeV.
For a given $\eta$, the light element abundances
are sensitive to the cosmic expansion rate $H$
at this epoch, which is given
by the Friedmann equation
$H^2 = 8\pi G \rho_{\rm rel} \sim g_* T^4/m_{\rm pl}^2$, 
and is sensitive (through $g_*$) to
the number of relativistic degrees of
freedom in equilibrium.  
Thus the observed primordial abundances
measure the number of relativistic species at the epoch of 
BBN, usually expressed in terms of the effective or
equivalent number of neutrino species $\nnu$
 \cite{ssg}.
By standard BBN we mean that
$\eta$ is homogeneous and the number of massless species of
neutrinos, $\nnu = 3$.
In this case, BBN has only one free parameter, $\eta$.
We will for now, however, relax the assumption of exactly
three light neutrino species.
In this case,
BBN becomes a two-parameter theory, with light
element abundance predictions
a function of $\eta$ and $\nnu$.

\begin{figure}
\begin{center}
\epsfig{file=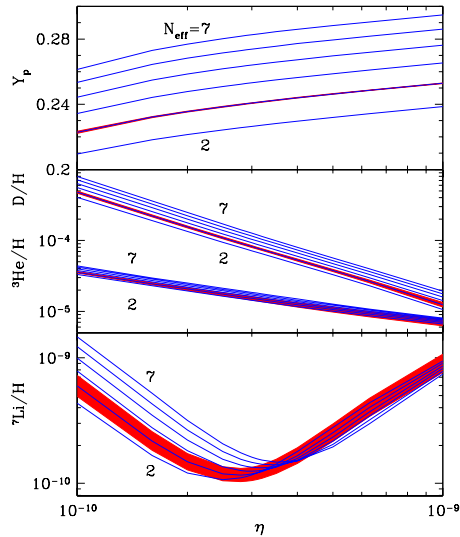,height=3.5in}
\end{center}
\vskip -.3in
\caption{
BBN abundance predictions as a function of the baryon-to-photon
ratio $\eta$, for $\nnu = 2$ to 7.  The bands show the $1\sigma$ error
bars.  Note that for the isotopes other than Li, the error bands are
comparable in width to the thickness of the abundance curve shown. All bands
are centered on $\nnu = 3$.}
\label{fig:etaNu}
\end{figure}

In Figure \ref{fig:etaNu}, we plot the primordial
abundances as a function of $\eta$ for a range
of \nnu\ from 2 to 7.   We see the usual offset in \he4,
but also note the shifts in the other elements, particularly
D, and also Li over some ranges in $\eta$.
Because of these variations, one is not restricted to only \he4 
in testing \nnu\ and 
particle physics. 

To quantify the predictions of BBN and their consistency
with the CMB, we 
adopt a likelihood analysis in the usual manner
 \cite{ot}.
Using BBN theory in Monte Carlo simulations,
one computes mean abundances,
usually quantified as 
$y_{{\rm th},i} = (Y_p,{\rm D/H},{\rm \he3/H},{\rm \li7/H})$,
and the theory error matrix $C_{ij}$
as functions of $\eta$  and $\nnu$.
Using these, one can construct a likelihood distribution
$\like_{\rm BBN}(\eta,\nnu;\vec{y})$.  One finds that the 
propagated errors are well approximated by gaussians,
in which case we can write
\beq
\label{eq:like-bbn}
\like_{\rm BBN}(\eta,\nnu;\vec{y}) =
  \frac{1}{\sqrt{(2\pi)^{N}|C|}} \
  \exp \left[  - \frac{1}{2}
        (\vec{y}-\vec{y}_{\rm th})^T C^{-1} (\vec{y}-\vec{y}_{\rm th}) 
       \right]
\eeq
This function
contains all of the statistical information
about abundance predictions and their correlations
at each $(\eta,\nnu)$ pair.

Eq. \pref{eq:like-bbn} can be used as follows:
\begin{enumerate}
\item {\em Testing BBN}: 
Typically, concordance is sought for the $\nnu =3$ case
\cite{wssok}.  Each value of the single free parameter $\eta$
predicts four light
nuclide abundances.
Thus the theory is overconstrained if
two or more primordial abundances are known.
With these abundances as inputs, it is possible to
determine if the theory is consistent with the data for
{\em any} range of $\eta$, and if so, one can 
determine the allowed $\eta$ range; this is the 
standard BBN prediction.

\item
{\em Probing the Early Universe}:
In this extension to case (1), one allows
for $\nnu \ne 3$, and uses two or more abundances
simultaneously to constrain $\eta$ and $\nnu$.
One therefore derives information about particle physics in the early
universe, via $\nnu$, as well as 
(somewhat looser) limits on $\eta$
\cite{ot,os95,other}.  This approach can be made quantitative by
convolving the likelihood in eq.\ \pref{eq:like-bbn} with an observational
likelihood function ${\cal L}_{\rm OBS}(\vec{y})$
\beq
\label{eq:OBS-BBN}
\like_{\rm OBS-BBN}(\eta,\nnu) 
 = \int {\rm d}\vec{y} \ \like_{\rm BBN}(\eta,\vec{y},\nnu) \
                     \like_{\rm OBS}(\vec{y})
\eeq

\item
{\em Predicting Light Element Abundances}.
Because the theory is overdetermined, one can
use eq. \pref{eq:like-bbn} as a way to combine
one set of abundances to determine $\eta$ (typically for
$\nnu = 3$) while simultaneously predicting the remaining
abundances.  This procedure is less common, but has
been used \cite{fo} to predict
Li depletion given a \he4 and D, or to predict D astration
given Li and \he4.  

\end{enumerate}

With the advent of
the CMB measurements of $\eta$,
we can take a different approach to BBN.
Namely, the CMB anisotropy measurements are strongly sensitive to
$\eta$ and thus independently measure this parameter.
The expected precision of 
{\em MAP} is $\sigma_\eta/\eta \la 10\%$
while {\em Planck} should improve this to $\la 3\%$ \cite{cmberr}.
In fact, the CMB anisotropies are also weakly sensitive\footnote{
This discussion applies to species with $m \ll 1$ eV.
If, for example, one or more neutrino species has
$m \sim 1$ eV, this can have a stronger impact on the
CMB \cite{lopez}.}
to the value of $\nnu$,
primarily via the early integrated Sachs-Wolfe effect.
Thus, the CMB measurements will
produce a likelihood of the form
$\like_{\rm CMB}(\eta,\nnu)$.
In practice, the CMB sensitivity to $\nnu$ is significantly
weaker than that of BBN.
The current CMB limits are 
$\nnu \la 17$ (95\% CL) \cite{han},
and are very sensitive to the assumed priors
\cite{kssw}.
Thus, to simplify the following discussion, we will ignore the
CMB dependence on $\nnu$. 
We thus write the CMB distribution as $\like_{\rm CMB}(\eta)$, and the
convolution of this with BBN theory
\beq
\label{eq:cmb-bbn}
\like_{\rm CMB-BBN}(\vec{y},\nnu) 
  = \int {\rm d}\eta \ \like_{\rm BBN}(\eta,\vec{y},\nnu) \
                     \like_{\rm CMB}(\eta)
\eeq
This expression gives the relative likelihoods of the
primordial abundances as a function of the
CMB-selected $\eta$.  This will select
the allowed ranges in the abundances and in \nnu,
and is 
the starting point for our analysis.

\begin{figure}
\begin{center}
\epsfig{file=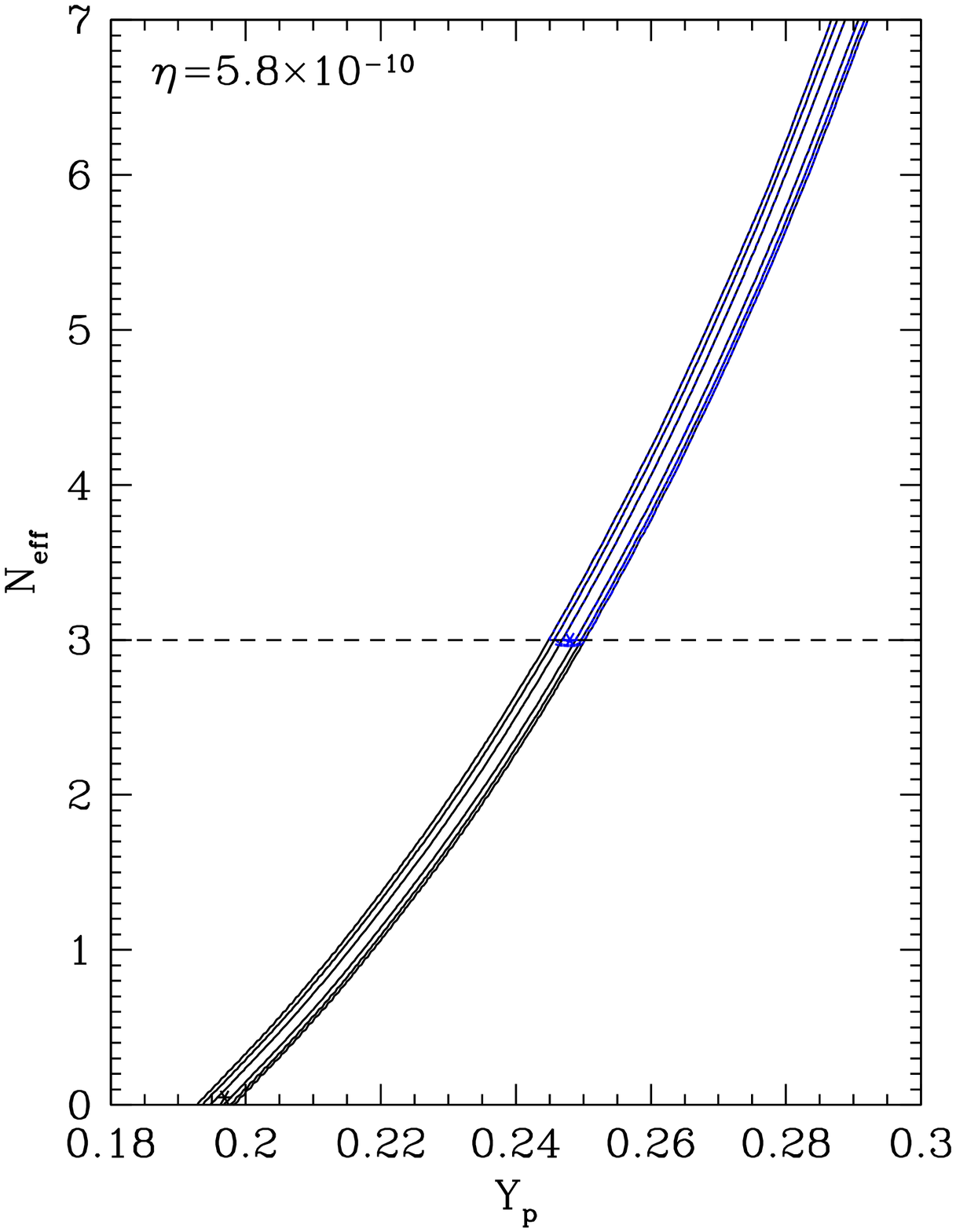,width=2in}
\epsfig{file=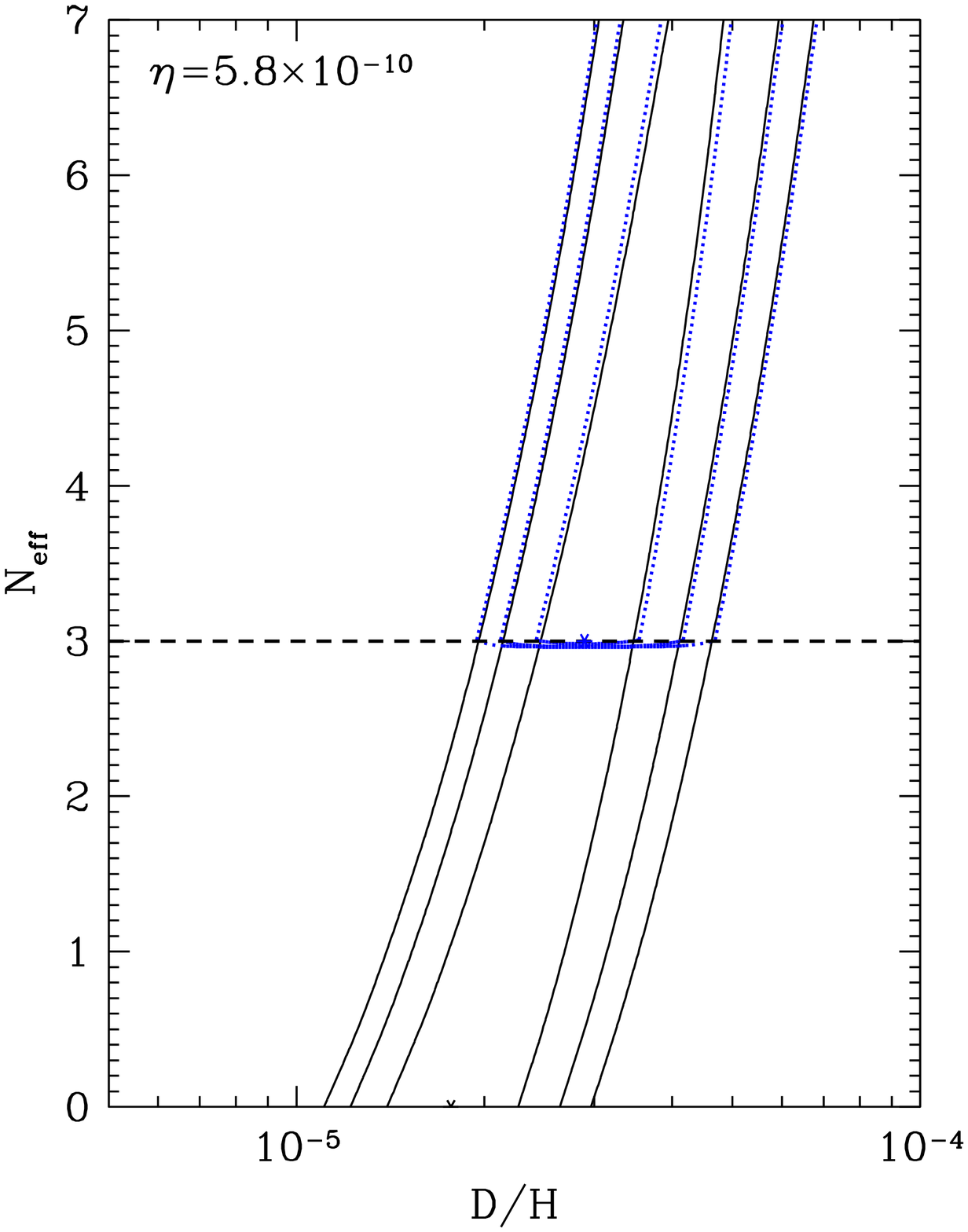,width=2in}
\epsfig{file=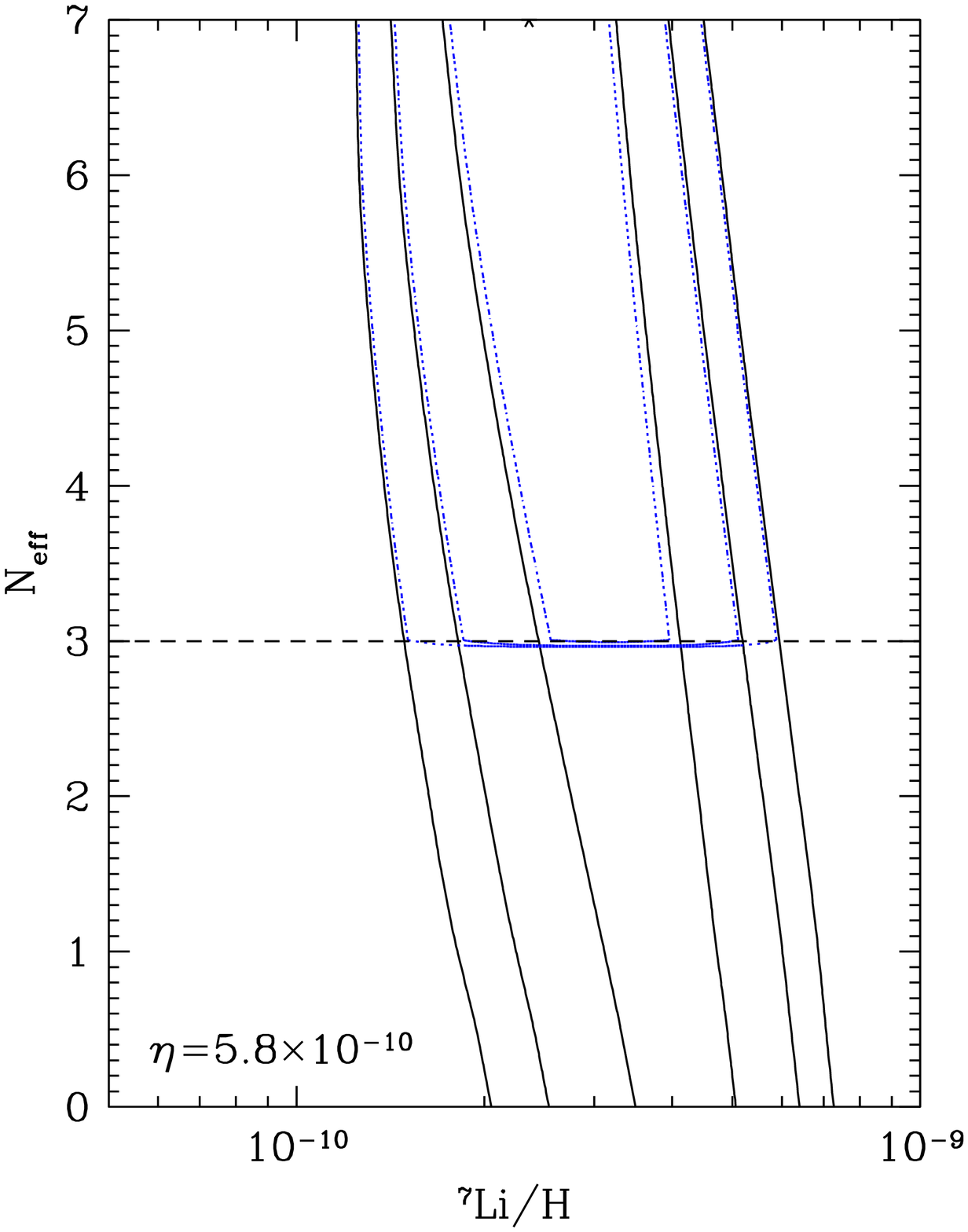,width=2in}
\end{center}
\vspace*{-.3in}
\caption{
The likelihood distribution of eq.\ \pref{eq:cmb-bbn} 
illustrated in its $y_i-\nnu$ projections.
Contours show 68\%, 95\%, and 99\% confidence levels,
and the peak likelihood is displayed as a star.
Solid curves are for the prior $0 \le \nnu \le 7$;
dotted curves are for $3 \le \nnu \le 7$.
We assume a CMB distribution in $\eta$ 
which is gaussian with $\eta_{10} = 5.8 \pm 0.58$,
i.e., the expected {\em MAP} error.
The departure from vertical in the peaks is a measure of the
ability to constrain $\nnu$; we see that this
is the strongest for \he4, but is possible to a lesser extent for
D and \li7.
\label{fig:cmb-eta1}
}
\end{figure}

Figure \ref{fig:cmb-eta1} illustrates
the combined likelihoods $\like_{\rm CMB-BBN}$ (projected on the $\eta - \nnu$
plane) one may expect using eq.\ \pref{eq:cmb-bbn} and assuming
a CMB determination of 
\beq
\label{eq:eta1}
\eta_{10} = 5.80 \pm 0.58 \ \ \mbox{(expected {\em MAP} error)}
\eeq
i.e., to a conservative 10\% accuracy,
based on the
``low'' deuterium observations \cite{omear}.
For simplicity we have used
a gaussian distribution, with a mean and standard deviation
as given in eq.\ \pref{eq:eta1}.  
For each element,
the likelihood forms a ``ridge'' in the
abundance--$\nnu$ plane, tracing the curve
$y_{i,{\rm max}}(\nnu) = y_i(\hat{\eta},\nnu)$ 
at the fixed $\hat{\eta}$ we have chosen.\footnote{
The peak likelihood value versus \nnu\ is
$\like(y_{i,{\rm max}},\nnu) = [\sqrt{2\pi}\sigma_i]^{-1}$,
and the slow variation of $\sigma_i(\hat{\eta},\nnu)$
leads to a small variation in the height of the ridge.
Thus, the maximum likelihood, denoted with a star,
falls at that point of the ridge corresponding to
the minimum in $\sigma_i(\hat{\eta},\nnu)$, typically
at the edge of the grid. 
However, as we see, the differences
in the height along the ridge are small, so that 
the CMB $\eta$, {\em by itself}, essentially serves to select
the $y_i-\nnu$ relation, and additional
information on one of these quantities then determines the other.
}
While the dependence on $\nnu$ is not dramatic
for any of the elements, the variation does exceed
the width of the ridge for
\he4, D, and \li7.  This sensitivity will
open the possibility for D and \li7 to probe $\nnu$.
We do not show \he3, as these contours appear as nearly vertical lines.
Note that a feature 
not apparent from these figures is the fact that the predicted
light element abundances are correlated for a given
$\nnu$; these correlations are essential to include
when combining information from predictions for
multiple elements.

\begin{figure}
\begin{center}
\epsfig{file=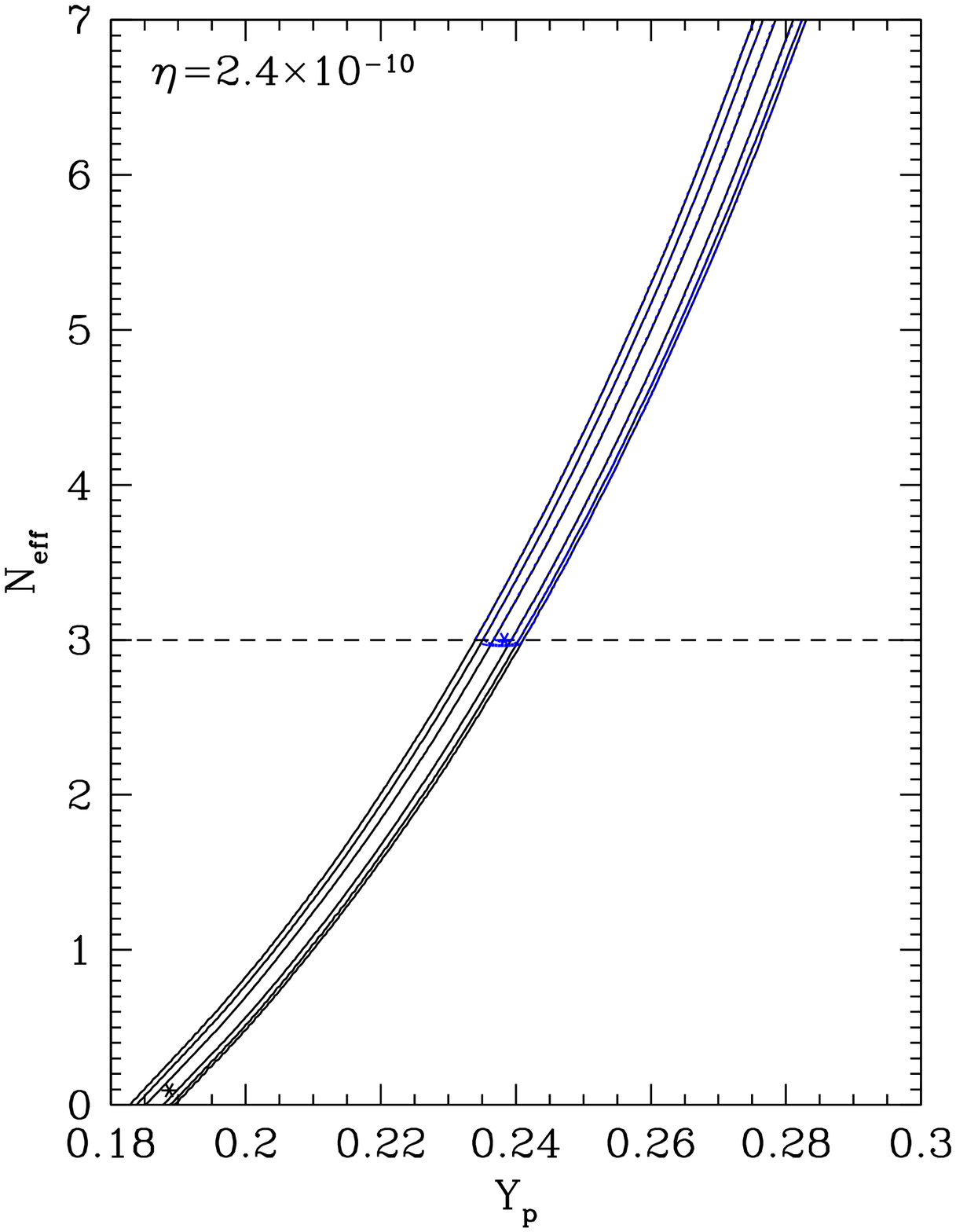,width=2in}
\epsfig{file=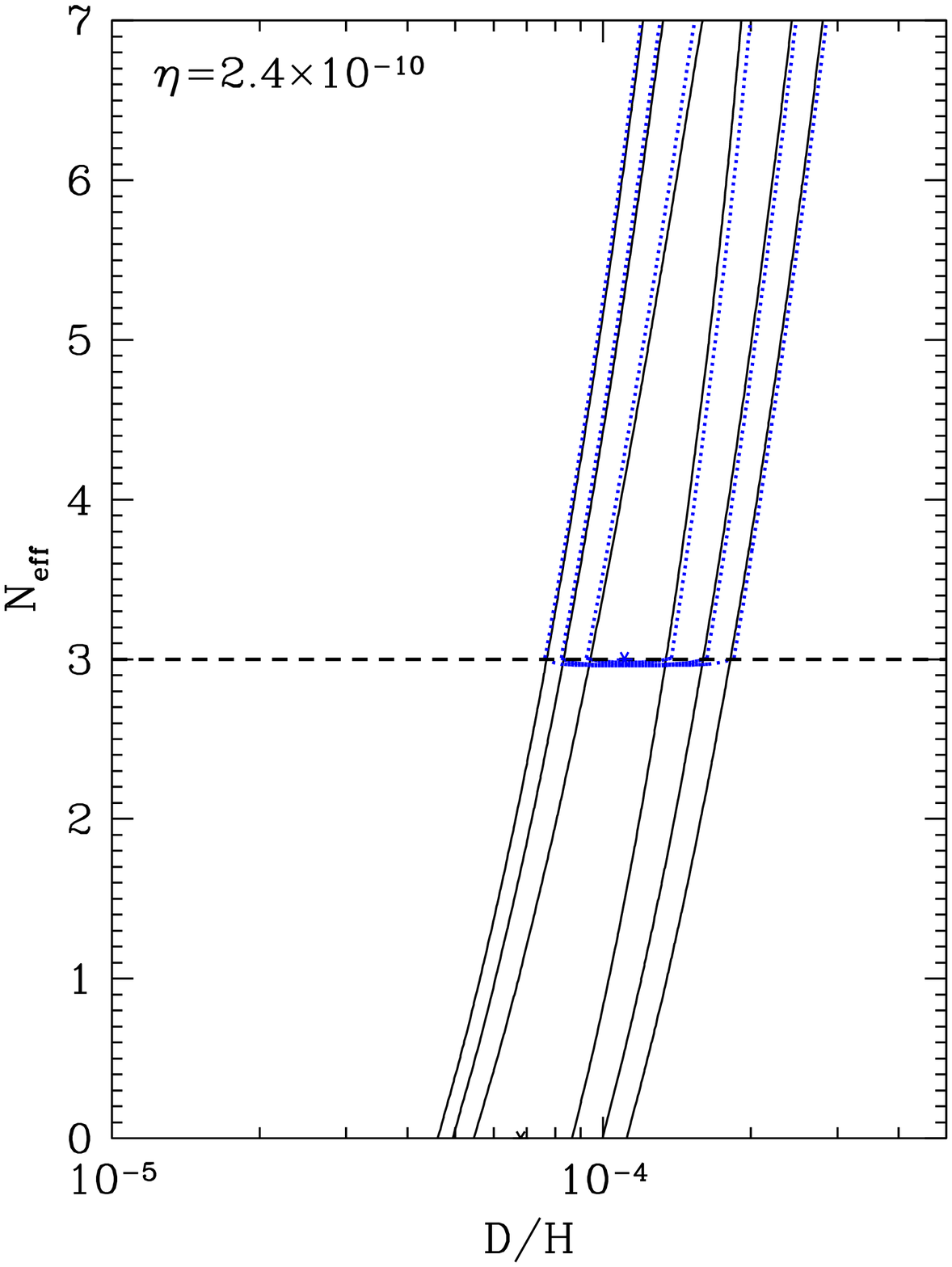,width=2in}
\epsfig{file=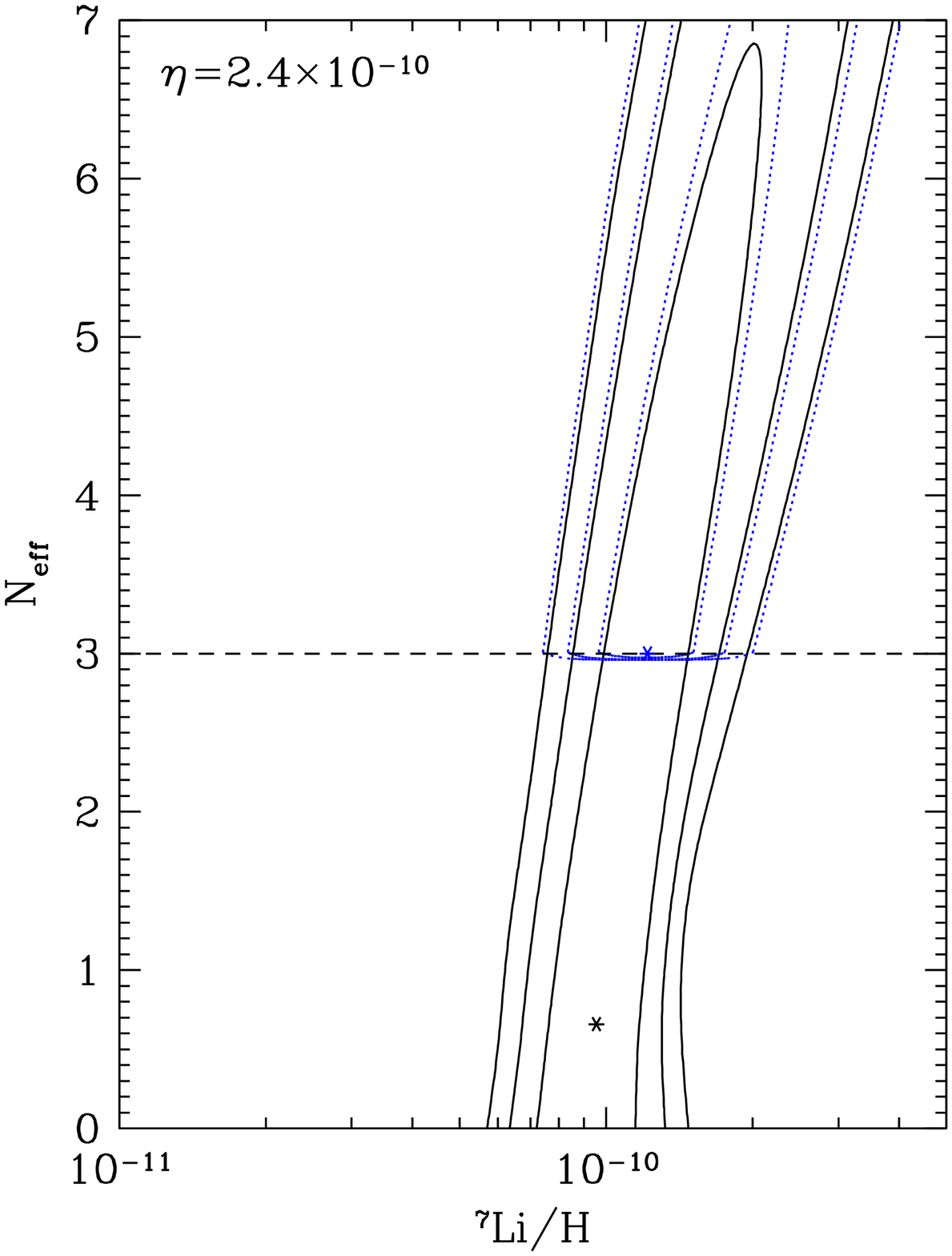,width=2in}
\end{center}
\vspace*{-.3in}
\caption{
As in figure \ref{fig:cmb-eta1}, with
${\eta}_{10} = 2.4 \pm 0.24$.
Note the change in sensitivity to $\nnu$.
\label{fig:cmb-eta2}
}
\end{figure}

The combined likelihood distribution of course
varies strongly with $\eta$, and so in
Figure \ref{fig:cmb-eta2} we illustrate
$\like_{\rm CMB-BBN}$ 
for
\beq
\label{eq:eta2}
\eta_{10} = 2.40 \pm 0.24 \ \  \mbox{(expected {\em MAP} error)}
\eeq
again, a 10\% measurement, this time
at the value favored by
\he4, \li7, and the higher D observation
\cite{cfo} (and by CBI \cite{cbi}).
Note that at this lower value of $\eta$,
D and Li show a slightly reduced
sensitivity to $\nnu$,
making
these elements somewhat weaker probes of this parameter.

\section{Testing BBN and Cosmology}
\label{sect:test}

The procedure to test BBN is conceptually simple, but 
the details of this crucial test are important.
For BBN, the difficulties 
stem from systematic uncertainties in 
the observational inference of abundances, and in 
the correction for post-BBN processing (chemical evolution) of
the light elements prior to the epoch at which they are observed.
Let us comment briefly on each of these in turn.

The \he4 data relevant for BBN comes from observations of \he4 in
low metallicity extragalactic H II regions.  
A correlation is found between the 
\he4 abundance and metallicity, and the primordial
abundance is extracted by extrapolating the available data to zero
metallicity. Because of the large number of very low metallicity
observations, this extrapolation is very sound statistically and yields
an error of only 0.002 (i.e. of only 1\%) in Y$_p$. However, the method
of analysis leads to a much larger uncertainty as can be seen by the
various results in the literature: $0.238 \pm 0.002$  \cite{oss}; $0.244 \pm 0.002$
 \cite{it}; $0.234 \pm 0.003$ \cite{ppr00}.  In addition, a recent detailed
examination of the systematic uncertainties in the \he4 abundance determination
showed that literature \he4 abundances typically under-estimated the true errors by
about a factor of 2 \cite{osk}.  The reason for the enhanced error determinations is
a degeneracy among the physical parameters  (electron density, optical depth, and 
underlying stellar absorption) which can yield equivalent results. 
Without new data or a reanalysis of the existing data, it is difficult
to ascribe a definite uncertainty to the \he4 abundance at this time.
For lack of a better number we will take our default value as
\beq
\label{eq:Y_p}
Y_p = 0.238 \pm 0.002 \pm 0.005
\eeq

As in the case of \he4, there is a considerable body of data on \li7 from
observations of hot halo dwarf stars.  Recent high
precision studies of Li abundances in halo stars have 
confirmed the existence of a plateau which signifies a primordial origin \cite{li}.
Ryan et al.\ \cite{rbofn} inferred a primordial Li abundance of 
\beq
\label{eq:Li_p}
\li7/{\rm H} = (1.23^{+0.68}_{-0.32}) \times 10^{-10}
\eeq
which includes a small correction for Galactic production which
{\em lowers} \li7/H compared to taking the mean value
over a range of metallicity.  
In contrast to the downward correction
due to post big bang production of Li, there is a potential for an upward
correction due to depletion.  Here, we note only that the data do not
show any dispersion (beyond that expected by observational uncertainty).  In this
event, there remains little room  for altering the \li7 abundance significantly.

The observational status of primordial D is very promising
if somewhat complicated.
Deuterium has been detected in several
high-redshift quasar absorption line systems.
It is expected that these systems still retain
their original,  primordial deuterium, unaffected by any significant
stellar nucleosynthesis.
At present, however, the determinations of D/H 
in different absorption systems show 
considerable scatter.  The result for D/H already used above
is \cite{omear}
\beq
{\rm D/H} = (3.0 \pm 0.4) \times 10^{-5}
\label{eq:lowd}
\eeq
and is based on three  determinations: 
D/H = $(3.3 \pm 0.3) \times 10^{-5}$, $(4.0 \pm 0.7) \times
10^{-5}$, and $(2.5 \pm 0.2) \times
10^{-5}$. 
O'Meara \etal\ \cite{omear} note, however, that $\chi^2_\nu = 7.1$ for 
the 3 combined D/H measurements (i.e., $\nu = 2$), 
and interpret this as a likely indication that the
errors have been underestimated.  There are in addition two other 
determinations: 
D/H = $(2.25 \pm 0.65) \times 10^{-5}$ \cite{dodo} and $(1.65 \pm 0.35) \times
10^{-5}$ \cite{pett}.  At the very least all of these measurements represents lower
limits to the primordial abundance.

On the CMB side, the key issue is the influence of a host
of parameters on
the anisotropy power spectrum.  That is, individual features in 
the power spectrum, such as peak heights and positions,
do depend on multiple parameters, and thus the measurement
of a few features can leave ambiguities in the inferred
cosmology.
Fortunately, different features in the power spectrum depend
differently on the cosmological parameters, so that
a sufficiently precise measurement with sufficient angular
coverage will be able to break the degeneracy represented by any
one feature.  Such precise measurements will be provided
by the space-based missions {\em MAP}
and {\em Planck}.
For the rest of the paper, we will assume the existence of
such measurements and thus an unambiguous determination of
$\like_{\rm CMB}(\eta)$, and examine the impact of
such a measurement on BBN.

\section{BBN After CMB Concordance}
\label{sect:BBN+CMB}

We now quantitatively explore the 
predictive power of BBN with $\eta$ given by the
CMB fluctuations.  The BBN predictions we use, and their
derivation from Monte Carlo calculations,
are described in detail in \cite{cfo}.

BBN neutrino counting will benefit significantly from 
the CMB revolution.
To date, BBN limits on \nnu\ require
an accurate \he4 abundance (to fix the neutrino number) and a good measure
of at least one more abundance (to fix $\eta$).
A precise determination of $\eta$ from the CMB anisotropy
opens up  other strategies.
One no longer need use light elements to fix $\eta$,
and thus all abundances are available to constrain \nnu.

Abundance observations fix the distributions
$\like_i^{\rm obs}(y_i)$.
We can convolve these with
eq.\ \ref{eq:cmb-bbn} to obtain
\beq
\like_{i \cdots \ell}(\nnu) = 
  \int {\rm d} y_i \, \cdots  \, {\rm d}y_\ell \
     \like_{\rm CMB-BBN}(\vec{y},\nnu) \
     \like_i(y_i) \cdots \like_\ell(y_\ell)
\eeq
a distribution for $\nnu$.
That is, for a given nuclide, the distribution in
$\nnu$ is given by the vertical region in 
Figure \ref{fig:cmb-eta1} or \ref{fig:cmb-eta2}
determined by the horizontal extent of the abundance measurement.
Of course, combining abundance determinations sharpens the limits
on $\nnu$.
One could then predict with great accuracy the abundances
for any elements not used for this analysis. 

\begin{figure}
\begin{center}
\epsfig{file=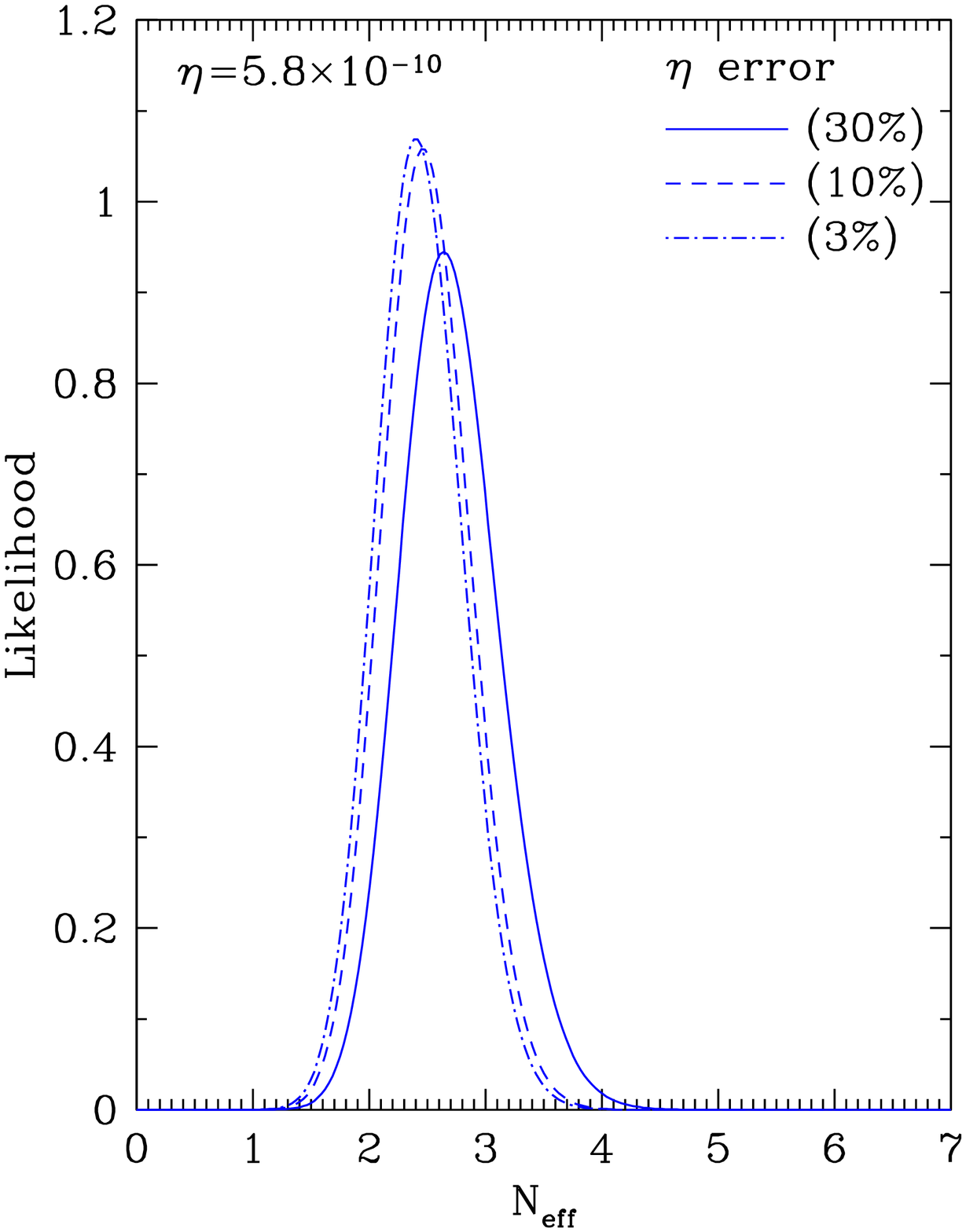,width=2in}
\epsfig{file=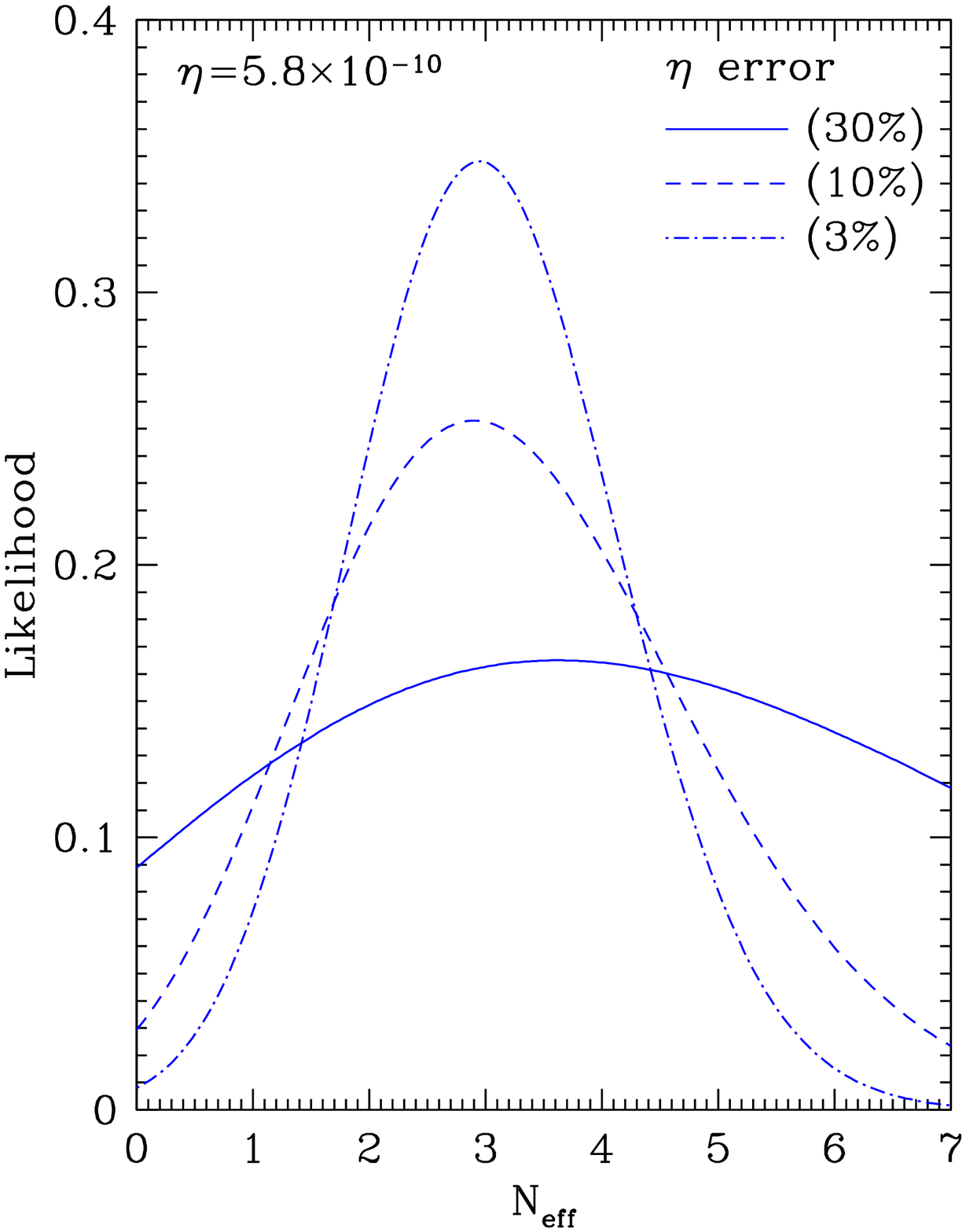,width=2in}
\epsfig{file=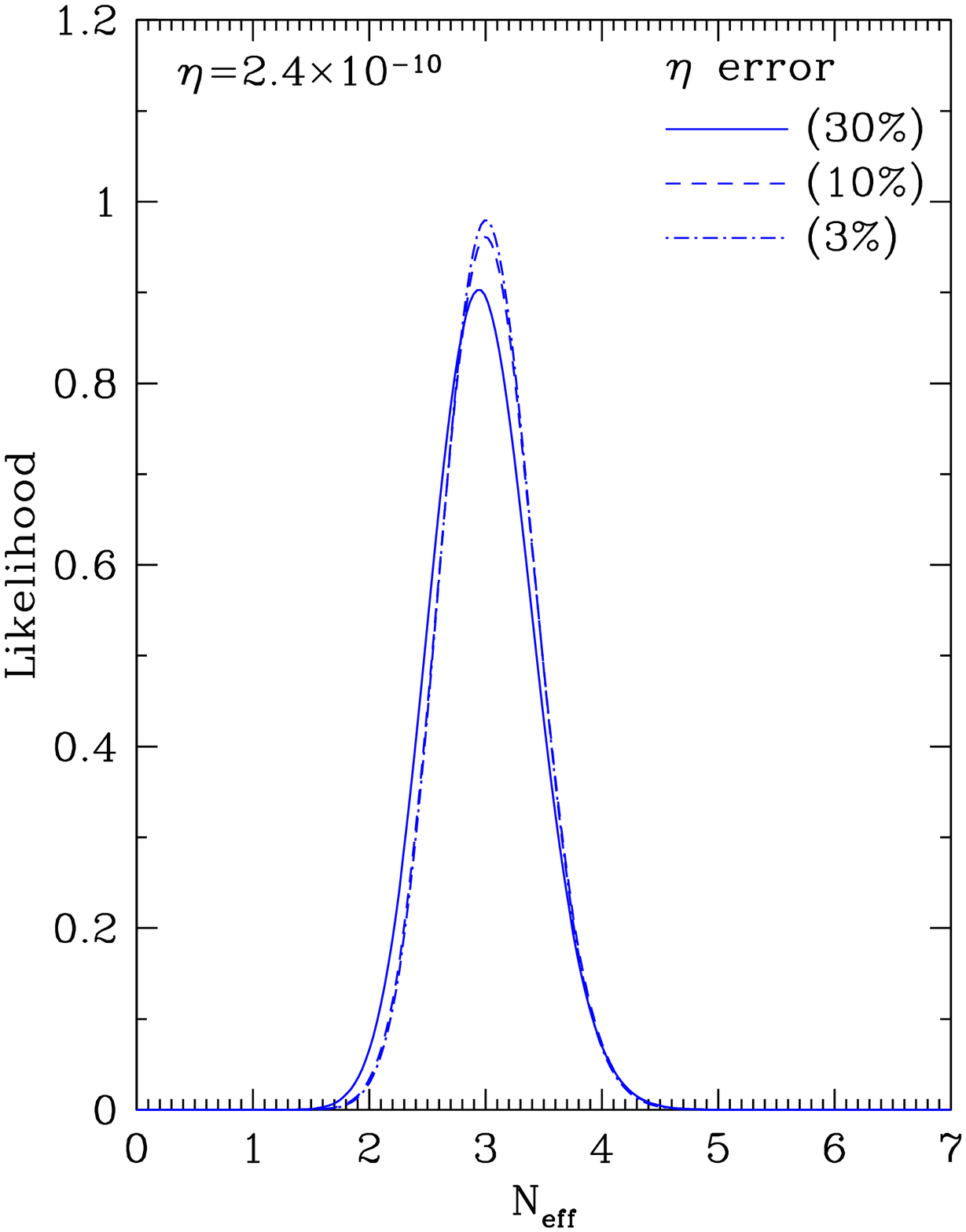,width=2in}
\end{center}
\vspace*{-.3in}
\caption{
(a)
The distribution in
\nnu\ assuming 
a CMB $\eta$ measurement of Figure \ref{fig:cmb-eta1},
and primordial \he4 \& \li7 abundances as
in eqs.\ \pref{eq:Y_p} \& \pref{eq:Li_p}.
The curves show the effect of the expected increased
accuracy in the CMB determination of $\eta$. (b) 
Distribution in
\nnu\ assuming 
a CMB $\eta$ measurement of Figure \ref{fig:cmb-eta1},
and a D measurement 
at the current precision (as in eq.\ \pref{eq:lowd}).
(c)  As in (a), but the 
a CMB $\eta$ measurement of Figure \ref{fig:cmb-eta2}.
\label{fig:Nnu-He}
}
\end{figure}

As expected, the \he4 abundance shows the most sensitivity to
$\nnu$.
Figure \ref{fig:Nnu-He}a illustrates the power of such an analysis in light
of CMB data. The curves show the resultant likelihood function for CMB measurements
of increasing accuracy (30, 10, and 3\%). If one can observe \he4\ to current
sensitivity, which we have assumed to be $\pm 0.0054$, we see that $\nnu$ can be
measured to a precision
$\sigma_{95\%}(\nnu) = 1.3$ (95\% CL) for a 30\% error in $\eta$, which
improves to $\sigma_{95\%}(\nnu) = 0.8$ for an uncertainty in $\eta$ of
$\le 10\%$. As one can see, with a CMB measurement of $\eta$ at even the 30\%
level, we are already be dominated by uncertainties in $Y_p$.
These limits, which depend only on \he4,
are comparable to present constraints \cite{ot}
which use BBN abundances to
fix the allowed value of $\eta$. But recall that the uncertainty in $Y_p$ may
actually be a factor of 2 larger \cite{osk}, in which case 
$\sigma_{95\%}(\nnu) = 1.7$ (95\% CL) for an uncertainty in $\eta$ of $\le
10\%$.

With $\eta$ independently and accurately fixed, it becomes
possible constrain $\nnu$ with nuclides other than, or in addition to,
\he4.   As seen in Figures \ref{fig:cmb-eta1} and \ref{fig:cmb-eta2},
deuterium shows a promising level of sensitivity to \nnu.
Indeed, D/H has been included in \nnu\ fitting by
several others
\cite{ot,os95,other},
although \he4 remained the primary
probe of \nnu. 
Figure \ref{fig:Nnu-He}b illustrates the predictive power 
of a measurement of 
${\rm D/H} = (3.0 \pm 0.4) \times 10^{-5}$, 
as found in recent high-redshift determinations (though
systematic uncertainties might lead to larger errors).  
We find that
it is possible to obtain a constraint on
$\nnu$ to an accuracy $\sigma_{95\%}(\nnu) = 2.2$
{\em with deuterium alone} 
for $\eta_{10}=5.8\times(1 \pm 0.03) $. While this constraint is at present
weak, it can be sharpened considerably with
improved D abundances (see below in \S \ref{sect:improve}).

The availability of elements other than \he4 for
neutrino counting has several advantages.
Observations of the different elements have very different systematics,
so that one can circumvent longstanding concerns about
\he4 observations by simply using other elements.
Also, the prospects for improvement in deuterium abundances
are better than for \he4.
For example, many new quasars will be found in the Sloan Survey
(see, e.g., \cite{sloanqso}), which will lead to a larger set of candidate D/H
systems.  There is thus reason for optimism
that systematics in the determination of primordial deuterium
will be sorted out by looking at a large sample.

Turning to the case of lithium, we
see from Figures \ref{fig:etaNu} through \ref{fig:cmb-eta2}
that the \li7 is almost insensitive to $\nnu$.
For realistic errors in the observed primordial Li abundance
($\ga 30\%$  \cite{rbofn})
one cannot expect to 
use this element as an \nnu\ discriminant.
This may even be a virtue, as the weak dependence 
of \li7 on 
$\nnu$ means that lithium remains a powerful cross-check
of the basic BBN concordance with the CMB,
independent of possible variations in \nnu.

Of course, the tightest constraints on $\nnu$
would come from combining all available abundances,
including \he4,
in order to exploit its sensitivity to $\nnu$.
Even if one chooses to be very conservative regarding
the uncertainties in the observed $Y_p$, 
depending on the accuracy of the observed D/H,
useful additional constraints can still flow from
conservative error bars ($\Delta Y_p \sim \pm 0.010$),
or from adopting upper or lower bounds to $Y_p$.

With $\eta_{\rm CMB}$ in hand, one can 
use BBN not only to probe the early universe, 
but also to accurately predict the light
element abundances and thus to probe
astrophysics.
Again, the starting point is the distribution in
abundances and $\nnu$ given by eq.\ \pref{eq:cmb-bbn}
and in Figures \ref{fig:cmb-eta1} and  \ref{fig:cmb-eta2}.
One must first address
the $\nnu$ dependence of the predictions.
For a conservative prediction of the abundances, one can simply
marginalize over all allowed $\nnu$, 
which implicitly assumes that all values in one's grid
are equally likely.  One might also
simply adopt the range determined from the invisible
width of $Z^0$ decay,
which presently gives $\nnu = 3.00 \pm 0.06$
 \cite{pdg}; this value is sufficiently accurate
that one may simply adopt $\nnu = 3$ in this case,
i.e., formally $\like_{\rm expt}(\nnu) = \delta(\nnu - 3)$.

\begin{figure}
\begin{center}
\epsfig{file=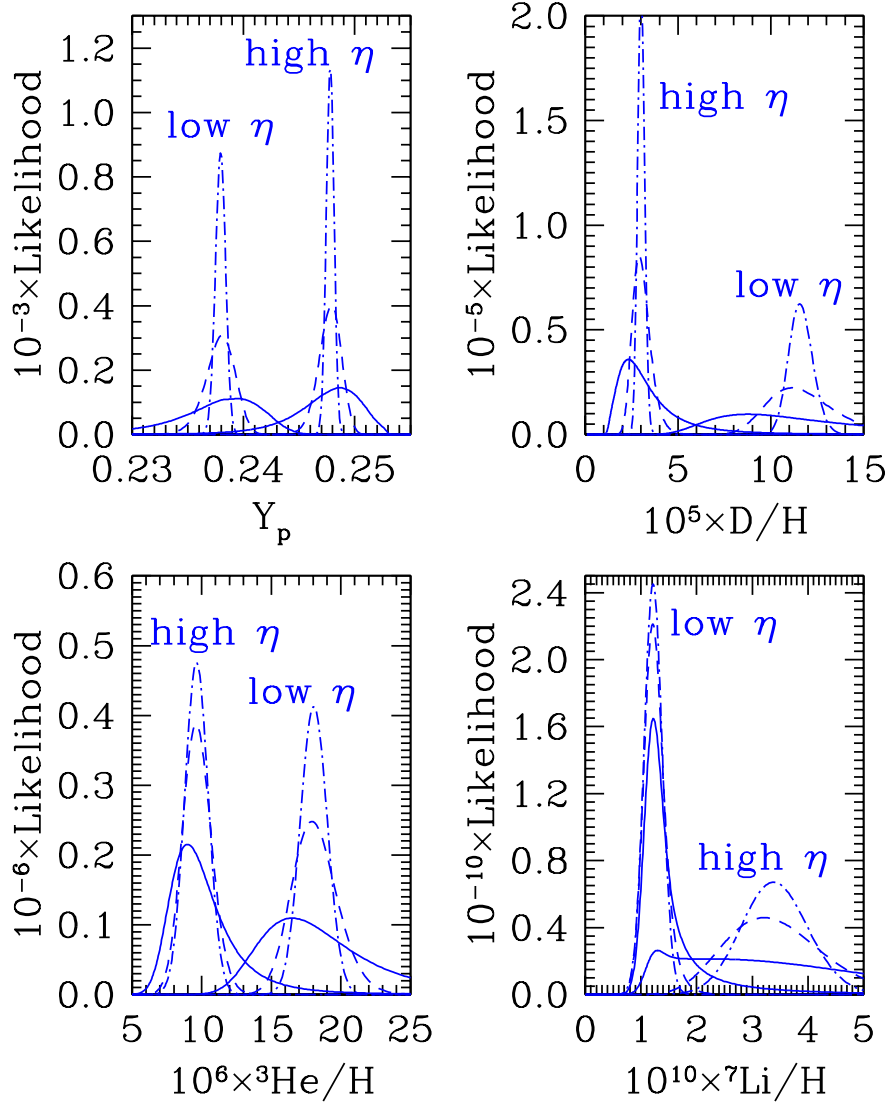,height=7in}
\end{center}
\caption{
Predicted abundances, illustrated
for two possible CMB-determined $\eta$ dubbed high(low) $\eta$
from eq.\ \ref{eq:eta1}(\ref{eq:eta2}).  Curves assuming (30, 10, 3\%)
error in $\eta$ are (solid, dashed, dot-dashed).
\label{fig:abs-eta}
}
\end{figure}

For the case $\nnu = 3$ we have computed the distribution of
predicted light element abundances.  
Results appear in Figure \ref{fig:abs-eta}.
One should bear in mind that for each $\eta$,
the light element predictions are
correlated, so that knowledge of one abundance will
narrow the distribution for the others.

With these predictions in hand, 
one can do astrophysics.
For example, deuterium is always destroyed astrophysically
\cite{els}.  Thus, 
the ${\rm D/D_p}$ ratio in
any astrophysical system is the
fraction of 
unprocessed material in that system, and
hence constrains the net amount of star formation. 
In the case of \he3, the stellar nucleosynthesis predictions
are uncertain, and the interpretation of dispersion
in the present-day observations is unclear;
a firm knowledge of the primordial abundance will 
provide a benchmark against which to infer the Galactic
production/destruction of \he3.
With regard to \li7, knowlege of the primordial abundance
can address issues of stellar depletion and will allow one
to better constrain the production of Li by early
Galactic cosmic rays \cite{rbofn}, and can
provide a consistency check on models of halo star
atmospheres.

\begin{figure}
\begin{center}
\epsfig{file=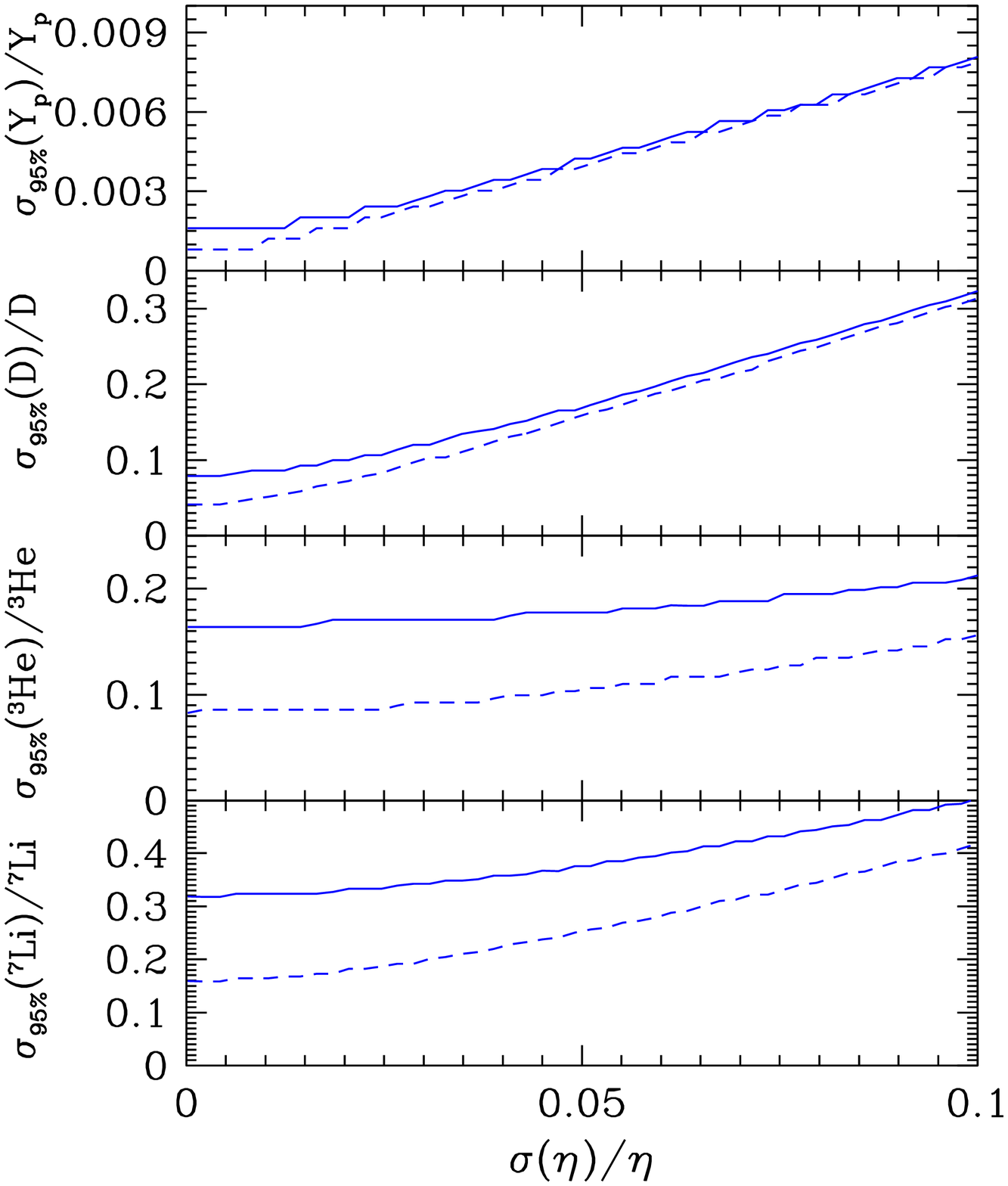,height=3.5in}
\vskip -.3in
\end{center}
\caption{
The 95\% CL accuracies of the abundance predictions
as a function of the CMB accuracy.  The adopted
$\eta$ is that of eq.\ \pref{eq:eta1}.
{\em Solid line:} present BBN theory errors.
{\em Dashed line:} BBN theory errors reduced by 50\%.
\label{fig:pred-acc}.
}
\end{figure}

Figure \ref{fig:pred-acc} illustrates the impact of
improvements in
the accuracy of the CMB $\eta$.
The solid curves show the fractional error (95\% CL)
$\sigma_{95\%}(y)/y$ in each element as a function of the CMB
precision.  We see that for \he4, D, and \li7,
the precision of the predictions can be improved significantly
with improved $\eta$ determinations.
This holds until $\sigma_\eta/\eta \simeq 0.7\%-3\%$
({\em Planck}'s expected level of
precision).  At this point,
the BBN theory errors begin to dominate;
we now turn to this issue.

\section{Needed Improvements in Observational and Theoretical Inputs}
\label{sect:improve}

\begin{figure}
\begin{center}
\epsfig{file=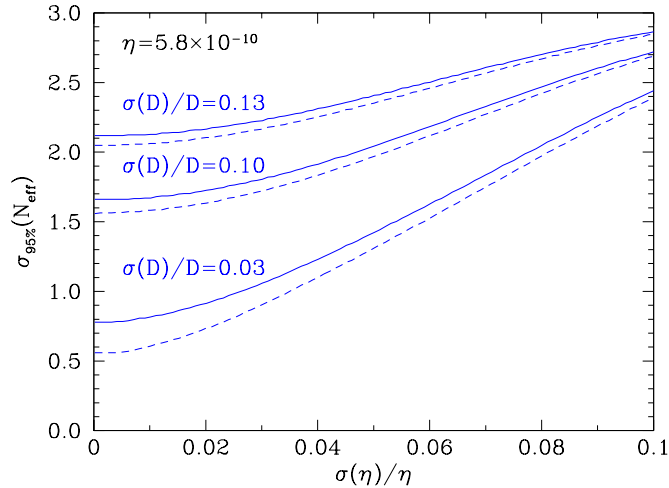,height=4in,angle=270}
\end{center}
\caption{
The precision of $\nnu$ determinations for different
levels of precision in D and $\eta$  measurements (95\% CL).
\label{fig:Nnu-Dacc}
}
\end{figure}

In anticipation of this new role for BBN,
it is important to note the limitations
to the power of BBN with $\eta_{\rm CMB}$ given.
The sharpness of the predictions is limited
by the precision of the observed primordial abundances,
and of the nuclear physics inputs.
On the observational side, as noted above we can
reasonably expect an improvement in D/H as
the number of high-redshift absorption line
systems increases.
To have an idea of the impact of lowering deuterium
observational errors,
we compute the $\nnu$ prediction using a
${\rm D/H} = 3.0 \times 10^{-5}$. 
We show in Figure \ref{fig:Nnu-Dacc} results with an uncertainty 
$\sigma({\rm D/H}) = 0.4 \times 10^{-5}$ as before,
i.e., $\sigma({\rm D})/{\rm D} = 13\%$, as well as  
$\sigma({\rm D})/{\rm D} = 10\%$
and 3\%.
We see that improved accuracy in the observed D abundances
considerably reduces the uncertainty in 
$\nnu$.  Even in the limit of a perfect
CMB observation ($\sigma_\eta/\eta \rightarrow 0$)
the observational and theoretical uncertainty in
D leads to a nonzero $\nnu$ width.
This reinforces the need to get an accurate determination of 
D/H as possible.  

The other source of uncertainty is the theoretical
error which stems from uncertainties in 
the nuclear inputs.
To show the effect of the current nuclear uncertainties,
we compute the predicted abundances by arbitrarily reducing the adopted errors
\cite{cfo} by 50\%.
These appear as the dashed curves in Figures
\ref{fig:pred-acc} and \ref{fig:Nnu-Dacc}.
We see that the theory 
uncertainties
are in fact a minor contributor to the total error budget of $\nnu$,
which is dominated by the observational abundance errors.
However, for abundance predictions,
the theory errors can be important, particularly for
\he3 and \li7.  Of these, the \li7 errors are the most
important candidate for improvement (e.g., \cite{cfo}), 
as an accurate knowledge of primordial \li7 can 
have an immediate impact on studies of Population II
stellar evolution, and on early Galactic cosmic rays.
On the other hand, our current theoretical
and observational understanding of \he3 is
probably too crude to profit from the high-precision
depicted in Fig.\ \ref{fig:pred-acc}, though this may
change by the time the {\em Planck} results are available.

\section{Discussion and Conclusions}
\label{sect:dis}

Upcoming precision measurements of 
CMB anisotropies will revolutionize almost all
aspects of cosmology.  
These data will allow for an independent and precise
measure of the baryon-to-photon ratio 
$\eta \propto \Omega_{\rm B} h^2$, and 
will thus have a major impact on BBN.
As we have shown,
even if there is agreement between CMB results and BBN predictions, BBN will not
lose its relevance for cosmology, but rather shifts its role and primary focus to
become a sharper probe of early universe particle physics and
of astrophysics. 

At present, the 20\% quoted uncertainty in $\eta$ from
CMB determinations \cite{boom,dasi,newboom} 
and the helium abundance of eq.\ \pref{eq:Y_p}
leads to the following
95  \% CL upper limits on \nnu:
\begin{itemize}
\item Using \he4 and $\eta_{10} = 5.8$ \qquad $\nnu < 3.6$
\item Using \he4 and $\eta_{10} = 2.4$ \qquad $\nnu < 3.9$
\end{itemize}
Note that in the first case, we have applied the prior that
$\nnu \ge 3.0$ \cite{os95},
while in the second
case we have shown the effect of a CMB $\eta$ consistent
with the \he4 and \li7 abundances and measured
to 20\%.\footnote{
In the first case, a prior of $\nnu \ge 0$ (2) leads
to a limit of $\nnu < 3.0$ (3.1).  In the second case,
a prior of $\nnu \ge 2$ (3) yields the limits
of $\nnu < 3.9$ (4.1).
}  
The 20\% uncertainty in $\eta$, in conjunction with
D/H as in eq.\ \pref{eq:lowd}
(i.e., with a 13\% uncertainty) 
gives a very weak limit on \nnu\ (i.e., the 95\%
CL limit is above $\nnu = 7$).
Future CMB determinations will tighten the \nnu\ bound.
While the limit based on \he4 is relatively unaffected by an
improved CMB determination (see Fig.\ \ref{fig:Nnu-He}), the
limit based on D will improve.
For example, with a 10\% measurement of $\eta$, and 
with D/H as in eq.\ \pref{eq:lowd}, 
the limit is $\nnu < 5.9$ (95\% CL).
With a 10\% (3\%) uncertainty in D/H,
the limit to \nnu\ is reduced to 5.7 (5.4).
Finally, the bound
will be reduced to $\nnu < 4.0$, assuming a
3\% uncertainty in both $\eta$ and D/H.

With $\eta_{\rm CMB}$ and light element abundances as
inputs, BBN will be able to better constrain
the physical conditions in the early universe.
We have illustrated this in terms of the effective number $N_{\nu,{\rm
eff}}$
of light neutrino species.  All light element abundances
will become available to constrain $\nnu$, allowing for
tighter limits and cross checks that are unavailable today.
We note in particularly the deuterium measurements alone
will be able to obtain useful limits on $\nnu$, independent 
of the use of \he4 observations.
Also, while we have framed the early universe physics
in terms of $\nnu$, one may also bring the power of the CMB
inputs to constrain a wide range of physics beyond the standard model
\cite{sarkar}, and more complicated 
early universe scenarios
such as inhomogeneous BBN \cite{ibbn}.

One can also use BBN theory and the CMB $\eta$ to infer
primordial abundances quite accurately.
This will sharpen our knowledge of astrophysics, 
with galactic-scale stellar processing probed via deuterium
abundances, and stellar nucleosynthesis
constrained with \he3 and \he4, 
and cosmic rays and stellar depletion tested with \li7.

In anticipation of the CMB results,
continued improvements in
light element observations and in BBN theory  are needed.
Reduced (but realistic!) error budgets 
are the key obstacle 
in maximizing leverage of the CMB $\eta$.
For light element observations, the key issue is
that of systematic errors.
For BBN theory, nuclear uncertainties now dominate
the error budget.
Efforts to improve both theory and observations will be
rewarded by the ability to do precision cosmology
with BBN.

\vskip 0.5in
\vbox{
\noindent{ {\bf Acknowledgments} } \\
\noindent  
The work of K.A.O. was partially supported by DOE grant
DE--FG02--94ER--40823.}


\begin{thebibliography}{99}

\bibitem{boom} 
P.~de Bernardis {\it et al.},
Nature {\bf 404}, 955 (2000)
[astro-ph/0004404];
A.~Balbi {\it et al.},
Astrophys.\ J.\  {\bf 545}, L1 (2000)
[astro-ph/0005124];
A. Jaffe, \etal\  Phys.\ Rev.\ Lett.\ {\bf 86}, 3475 (2000) [astro-ph/0007333].

\bibitem{cbi} S. Padin, S., et al.\ Astrophys.\ J.\  {\bf 549},
L1 (2001).

\bibitem{dasi} C. Pryke \etal, (2001) [astro-ph/0104490].

\bibitem{newboom} C.B. Netterfield \etal, (2001) [astro-ph/0104460].

\bibitem{cmbtest}
D.~N.~Schramm and M.~S.~Turner,
Rev.\ Mod.\ Phys.\  {\bf 70}, 303 (1998)
[astro-ph/9706069];
S.~Burles, K.~M.~Nollett and M.~S.~Turner,
Phys.\ Rev.\ D {\bf 63}, 063512 (2001)
[astro-ph/0008495].

\bibitem{omear} J.M. O'Meara, \etal, (2000)
[astro-ph/0011179].

\bibitem{cfo} R.H. Cyburt, B.D. Fields, and K.A. Olive, 
New Astron., in press (2001) [astro-ph/0102179].

\bibitem{ssg} G. Steigman, D.N. Schramm, and J. Gunn, Phys.\ Lett.\ {\bf B66}, 202
(1977).

\bibitem{ot}
K.~A.~Olive and D.~Thomas,
Astropart.\ Phys.\  {\bf 7}, 27 (1997)
[hep-ph/9610319];
K.~A.~Olive and D.~Thomas,
Astropart.\ Phys.\  {\bf 11}, 403 (1999)
[hep-ph/9811444];
E.~Lisi, S.~Sarkar and F.~L.~Villante,
Phys.\ Rev.\ D {\bf 59}, 123520 (1999)
[hep-ph/9901404].

\bibitem{wssok}
T.~P.~Walker, G.~Steigman, D.~N.~Schramm, K.~A.~Olive and H.~Kang,
Astrophys.\ J.\  {\bf 376}, 51 (1991);
K.~A.~Olive, G.~Steigman and T.~P.~Walker,
Phys.\ Rept.\  {\bf 333}, 389 (2000)
[astro-ph/9905320].

\bibitem{os95}
K.~A.~Olive and G.~Steigman,
Phys.\ Lett.\ B {\bf 354}, 357 (1995)
[hep-ph/9502400].

\bibitem{other}
C.~Y.~Cardall and G.~M.~Fuller,
Astrophys.\ J.\  {\bf 472}, 435 (1996)
[astro-ph/9603071];
P.~J.~Kernan and S.~Sarkar,
Phys.\ Rev.\ D {\bf 54}, 3681 (1996)
[astro-ph/9603045]
N.~Hata, G.~Steigman, S.~Bludman and P.~Langacker,
Phys.\ Rev.\ D {\bf 55}, 540 (1997)
[astro-ph/9603087];
C.~J.~Copi, D.~N.~Schramm and M.~S.~Turner,
Phys.\ Rev.\ D {\bf 55}, 3389 (1997)
[astro-ph/9606059].

\bibitem{fo}
B.~D.~Fields and K.~A.~Olive,
Phys.\ Lett.\ B {\bf 368}, 103 (1996)
[hep-ph/9508344].




\bibitem{cmberr}
M.~Zaldarriaga, D.~N.~Spergel and U.~Seljak,
Astrophys.\ J.\  {\bf 488}, 1 (1997)
[astro-ph/9702157];
J.~R.~Bond, G.~Efstathiou and M.~Tegmark,
Monthly Not.\ Royal Astr.\ Soc.\ {\bf 291}, L33 (1997)
[astro-ph/9702100];
G.~Jungman, M.~Kamionkowski, A.~Kosowsky and D.~N.~Spergel,
Phys.\ Rev.\ D {\bf 54}, 1332 (1996)
[astro-ph/9512139].

\bibitem{lopez}
R.~E.~Lopez,
(1999) [astro-ph/9909414].

\bibitem{han}
S.~Hannestad,
Phys.\ Rev.\ Lett.\  {\bf 85}, 4203 (2000)
[astro-ph/0005018].
S.~Hannestad,
``New CMBR data and the cosmic neutrino background,''
[astro-ph/0105220].

\bibitem{kssw}
J.~P.~Kneller, R.~J.~Scherrer, G.~Steigman and T.~P.~Walker,
astro-ph/0101386.


\bibitem{oss}
K.~A.~Olive, E.~Skillman and G.~Steigman,
astro-ph/9611166;
B.~D.~Fields and K.~A.~Olive,
Astrophys.\ J.\  {\bf 506}, 177 (1998)
astro-ph/9803297.

\bibitem{it}
Y.I. Izotov, and T.X. Thuan,  Astrophys.\ J.\  {\bf 500}, 188 (1998).

\bibitem{ppr00}
M. Peimbert, A. Peimbert, and M.T. Ruiz, M. T. Astrophys.\ J.\  {\bf 541},
688 (2000).

\bibitem{osk} K.A. Olive, and E. Skillman, E.  New Ast.,
in press, (2001) [astro-ph/0007081].

\bibitem{li}
P. Bonifacio, P. and P. Molaro, P., MNRAS {\bf 285},
847 (1997);
S.G. Ryan, J. Norris, and T.C. Beers,  Astrophys.\ J.\  {\bf 
523}, 654 (1999).

\bibitem{rbofn}
S.~G.~Ryan, T.~C.~Beers, K.~A.~Olive, B.~D.~Fields and J.~E.~Norris,
Astrophys.\ J.\  {\bf 530}, L57 (2000)
[astro-ph/9905211].

\bibitem{dodo}
S. D'Odorico, M. Dessauges-Zavadsky, and P. Molaro, Ast.\ Astro.\
(2001) [astro-ph/0102162].

\bibitem{pett}
M. Pettini and D.V. Bowen, (2001) astro-ph/0104474.


\bibitem{sloanqso}
D.~P.~Schneider {\it et al.}\ Astronom.~J. {\bf 121}, 1232 (2001) 

\bibitem{els}
R. Epstein, J. Lattimer and D.N. Schramm, Nature {\bf 263}, 198 (1976). 

\bibitem{pdg}
D.~E.~Groom {\it et al.}  [Particle Data Group Collaboration],
Eur.\ Phys.\ J.\ C {\bf 15}, 1 (2000).

\bibitem{sarkar}
S.~Sarkar,
Rept.\ Prog.\ Phys.\  {\bf 59}, 1493 (1996)
[hep-ph/9602260].

\bibitem{ibbn}
H.~Kurki-Suonio and E.~Sihvola,
Phys.\ Rev.\ D {\bf 63}, 083508 (2001)
[astro-ph/0011544];
K.~Jedamzik, G.~M.~Fuller and G.~J.~Mathews,
Astrophys.\ J.\  {\bf 423}, 50 (1994)
[astro-ph/9312065].

\end{thebibliography}
\end{document}